\documentclass[conference,a4paper]{IEEEtran}
\usepackage{geometry}                % See geometry.pdf to learn the layout options. There are lots.
\usepackage{graphicx}
\usepackage{amssymb}
\usepackage{amsmath}
\usepackage{epstopdf}
\usepackage{subfigure}
\usepackage{amsthm}
\newtheorem*{mydef}{Definition}
\title{Determinism, Complexity, and Predictability in Computer Performance}

 \author{
   \IEEEauthorblockN{
     Joshua Garland\IEEEauthorrefmark{1},
     Ryan G.~James\IEEEauthorrefmark{3} and
     Elizabeth Bradley\IEEEauthorrefmark{1}\IEEEauthorrefmark{2}}
   \IEEEauthorblockA{
     \IEEEauthorrefmark{1}Dept. of Computer Science
     University of Colorado, Boulder, Colorado 80309-0430 USA\\
     Email: joshua.garland@colorado.edu}
   \IEEEauthorblockA{
     \IEEEauthorrefmark{2}Santa Fe Institute, 1399 Hyde Park Road, Santa Fe, New Mexico 87501  USA\\
     Email: lizb@colorado.edu}
   \IEEEauthorblockA{
     \IEEEauthorrefmark{3}Complexity Sciences Center \& Physics Dept., University of California, Davis, California 95616 USA\\
     Email: rgjames@ucdavis.edu}
 }

\begin{document}

\maketitle

\begin{abstract}
Computers are deterministic dynamical systems \cite{mytkowicz09}.  Among other things, that implies that
one should be able to use deterministic forecast rules
to predict their behavior.  That statement is sometimes---but not
always---true.  The memory and processor loads of some simple programs
are easy to predict, for example, but those of more-complex programs
like {\tt gcc} are not.
%%%%%%%%%%%%%%%%%%%%%%
%% I had to change all the \verb|blah| entries because they caused
%% latex to barf if they were in figure captions.  Odd bug.
%%%%%%%%%%%%%%%%%%%%%%
The goal of this paper is to determine why that is the case.  We
conjecture that, in practice, complexity can effectively overwhelm the
predictive power of deterministic forecast models.  To explore that,
we build models of a number of performance traces from different
programs running on different Intel-based computers.  We then
calculate the \emph{permutation entropy}---a temporal entropy metric
that uses ordinal analysis---of those traces and correlate those values
against the prediction success.
\end{abstract}

\section{Introduction}
%\begin{it}
%Paragraph on computer performance, including citations to Todd paper
%and summary of the results that indicate that they're deterministic
%nonlinear dynamical systems.  Given that, we should be able to
%predict.  What benefits would accrue if we could do so: power mgmt,
%end world hunger [[this is my primary goal everyday :)]], etc.
%\end{it}

\begin{figure}[htbp]
   \centering
   \includegraphics[width=0.5\textwidth]{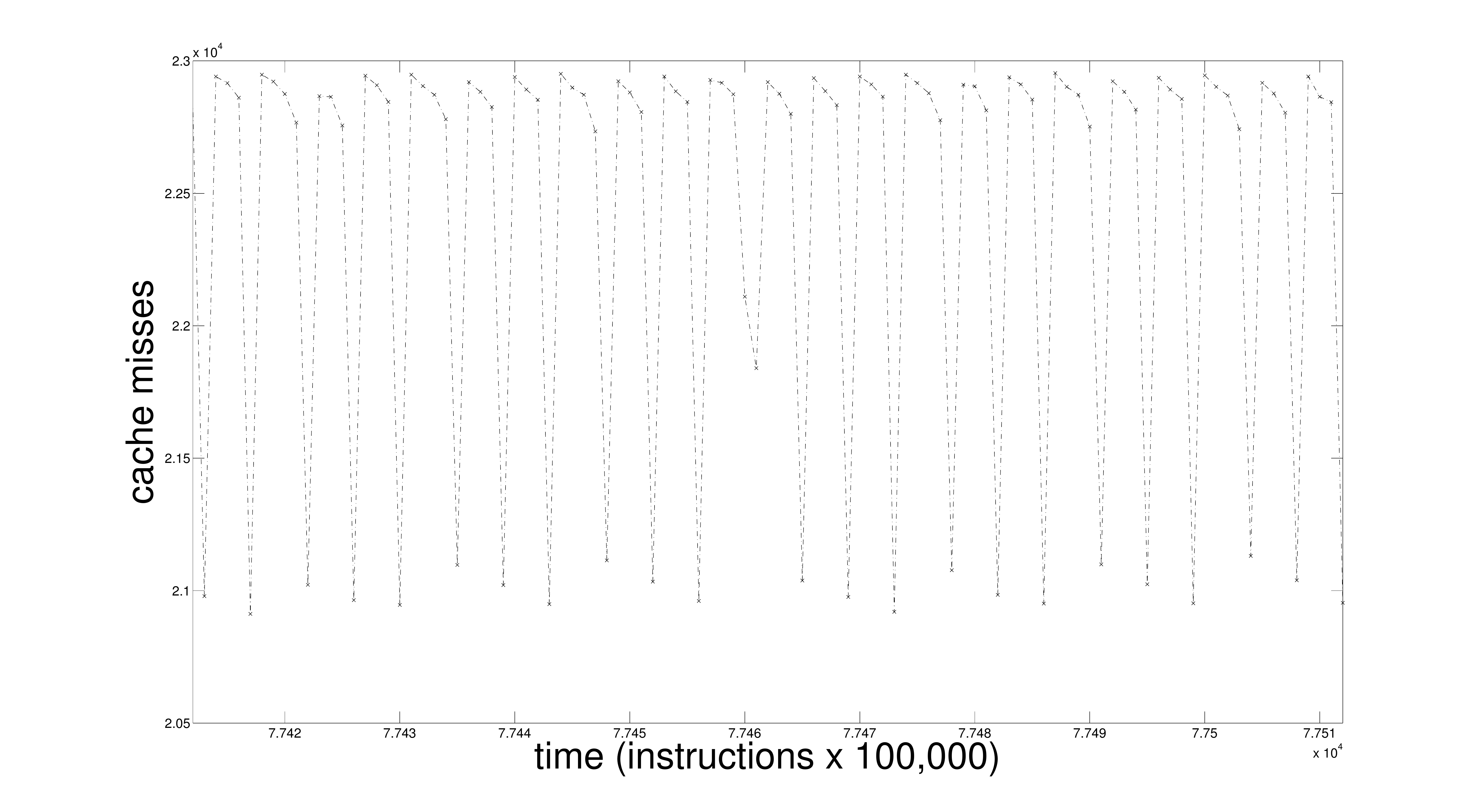}
   % where an .eps filename suffix will be assumed under latex,
   % and a .pdf suffix will be assumed for pdflatex
   \caption{A small snippet of the L2 cache miss rate of {\tt
       col\_major}, a three-line C program that repeatedly initializes
     a matrix in column-major order, running on an Intel Core
     Duo\textsuperscript{\textregistered}-based machine.  Even this
     simple program exhibits chaotic performance dynamics.}
   \label{fig:cache}
 \end{figure}

Computers are among the most complex engineered artifacts in current
use.  Modern microprocessor chips contain multiple processing units
and multi-layer memories, for instance, and they use complicated
hardware/software strategies to move data and threads of computation
across those resources.  These features---along with all the others
that go into the design of these chips---make the patterns of their
processor loads and memory accesses highly complex and hard to
predict.  Accurate forecasts of these quantities, if one could
construct them, could be used to improve computer design.  If one
could predict that a particular computational thread would be bogged
down for the next 0.6 seconds waiting for data from main memory, for
instance, one could save power by putting that thread on hold for that
time period (e.g., by migrating it to a processing unit whose clock
speed is scaled back).  Computer performance traces are, however, very
complex.  Even a simple ``microkernel,'' like a three-line loop that
repeatedly initializes a matrix in column-major order, can produce
{\sl chaotic} performance traces \cite{mytkowicz09}, as shown in
Figure~\ref{fig:cache}, and chaos places fundamental limits on
predictability.

The computer systems community has applied a variety of prediction
strategies to traces like this, most of which employ regression.  An
appealing alternative builds on the recently established fact that
computers can be effectively modeled as deterministic nonlinear
dynamical systems \cite{mytkowicz09}.  This result implies the
existence of a deterministic forecast rule for those dynamics.  In
particular, one can use \emph{delay-coordinate embedding} to
reconstruct the underlying dynamics of computer performance, then use
the resulting model to forecast the future values of computer
performance metrics such as memory or processor loads
\cite{josh-ida2011}.  In the case of simple microkernels like the one
that produced the trace in Figure~\ref{fig:cache}, this deterministic
modeling and forecast strategy works very well.  In more-complicated
programs, however, such as speech recognition software or compilers,
this forecast strategy---as well as the traditional methods---break
down quickly.

This paper is a first step in understanding when, why, and how
deterministic forecast strategies fail when they are applied to
deterministic systems.  We focus here on the specific example of
computer performance.  We conjecture that the complexity of traces
from these systems---which results from the inherent dimension,
nonlinearity, and nonstationarity of the dynamics, as well as from
measurement issues like noise, aggregation, and finite data
length---can make those deterministic signals \emph{effectively}
unpredictable.  We argue that \emph{permutation entropy}
\cite{bandt2002per}, a method for measuring the entropy of a
real-valued-finite-length time series through ordinal analysis, is an
effective way to explore that conjecture.  We study four
examples---two simple microkernels and two complex programs from the
SPEC benchmark suite---running on different Intel-based machines.  For
each program, we calculate the permutation entropy of the processor
load (instructions per cycle) and memory-use efficiency (cache-miss
rates), then compare that to the prediction accuracy attainable for
that trace using a simple deterministic model.

% The rest of the paper is organized as follows.
% Section~\ref{sec:compModel} describes the experimental setup, as well
% as the nonlinear modeling and forecast strategies.  In
% Section~\ref{sec:meaComplex}, we review permutation entropy, calculate
% its value for a number of different computer performance traces, and
% compare the results to the prediction accuracy.  In
% Section~\ref{sec:conc}, we discuss these results and their
% implications in regard to our conjecture, and consider future areas of
% research.

\section{Modeling Computer Performance}\label{sec:compModel}

% took out for space
% \subsection{Reconstructing hidden dynamics}

Delay-coordinate embedding allows one to reconstruct a system's full
state-space dynamics from a \emph{single} scalar time-series
measurement---provided that some conditions hold regarding that data.
Specifically, if the underlying dynamics and the measurement
function---the mapping from the unknown state vector $\vec{X}$ to the
scalar value $x$ that one is measuring---are both smooth and generic,
Takens~\cite{takens} formally proves that the delay-coordinate map
\[
F(\tau,m)(x) = ([x(t) ~ x(t+\tau) ~ \dots ~x(t+m\tau)])
\]
from a $d$-dimensional smooth compact manifold $M$ to ${Re}^{2d+1}$,
where $t$ is time, is a diffeomorphism on $M$---in other words, that
the reconstructed dynamics and the true (hidden) dynamics have the
same topology.

\begin{figure}
  \centering
    \includegraphics[width=0.5\textwidth]{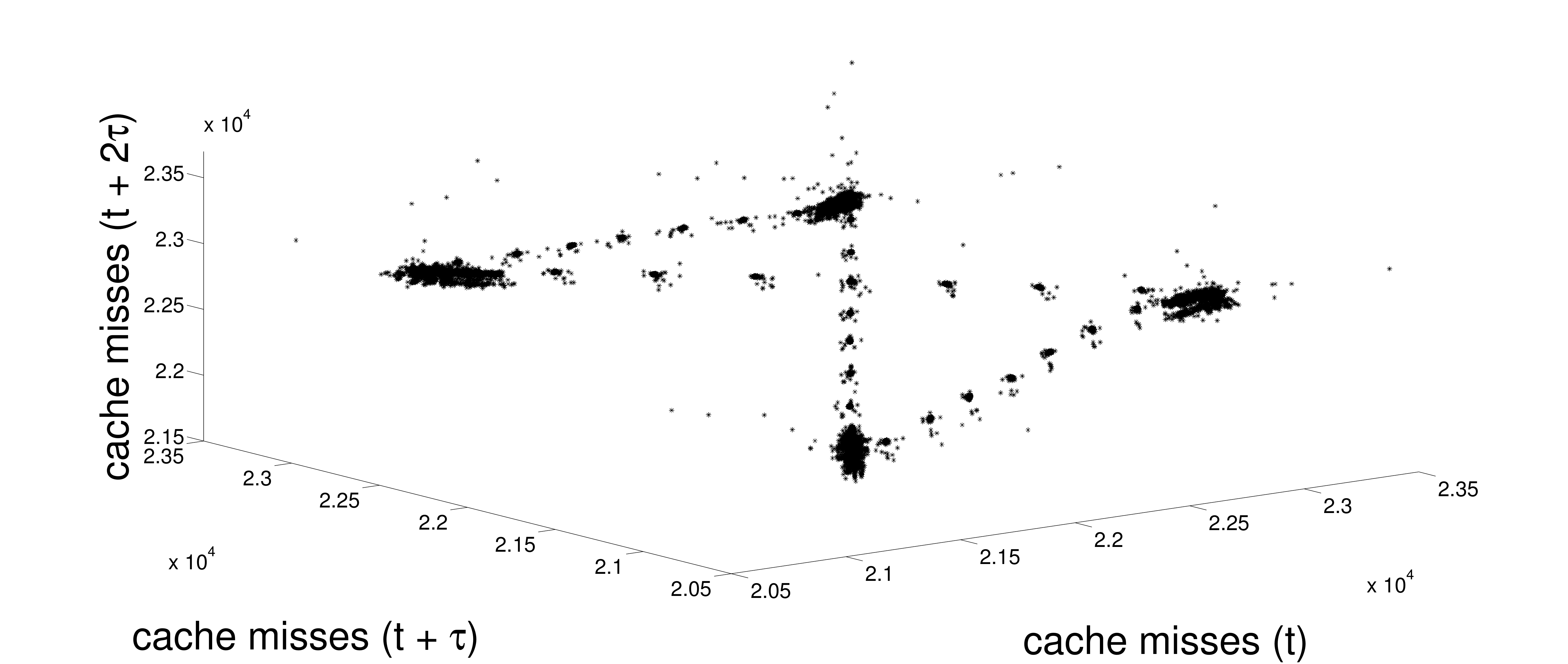}
    \caption{A 3D projection of a delay-coordinate embedding of the trace
from Figure~\ref{fig:cache} with a delay ($\tau$) of 100,000 instructions.
}
\label{fig:embedding}
\end{figure}

This is an extremely powerful result: among other things, it means
that one can build a formal model of the full system dynamics without
measuring (or even knowing) every one of its state variables.  This is
the foundation of the modeling approach that is used in this paper.
The first step in the process is to estimate values for the two free
parameters in the delay-coordinate map: the delay $\tau$ and the
dimension $m$.  We follow standard procedures for this, choosing the
first minimum in the average mutual information as an estimate of
$\tau$ \cite{fraser-swinney} and using the false-near(est) neighbor
method of \cite{KBA92}, with a threshold of 10\%, to estimate $m$.  A
plot of the data from Figure~\ref{fig:cache}, embedded following this
procedure, is shown in Figure~\ref{fig:embedding}.
%% can cut for space if need be:
The coordinates of each point on this plot are differently delayed
elements of the \verb|col_major| L2 cache miss rate time series
$y(t)$: that is, $y(t)$ on the first axis, $y(t+\tau)$ on the second,
$y(t+2\tau)$ on the third, and so on.
%% ...down to here.
Structure in these kinds of plots---clearly visible in
Figure~\ref{fig:embedding}---is an indication of
determinism\footnote{A deeper analysis of
  Figure~\ref{fig:embedding}---as alluded to on the previous
  page---supports that diagnosis, confirming the presence of a chaotic
  attractor in these cache-miss dynamics, with largest Lyapunov
  exponent $\lambda_1 = 8000 \pm 200$ instructions, embedded in a
  12-dimensional reconstruction space \cite{mytkowicz09}.}.  That
structure can also be used to build a forecast model.

% took out for space
% \subsection{LMA: Using dynamics in forecasting}

\begin{figure}[htbp]
  \centering
    \includegraphics[width=0.5\textwidth]{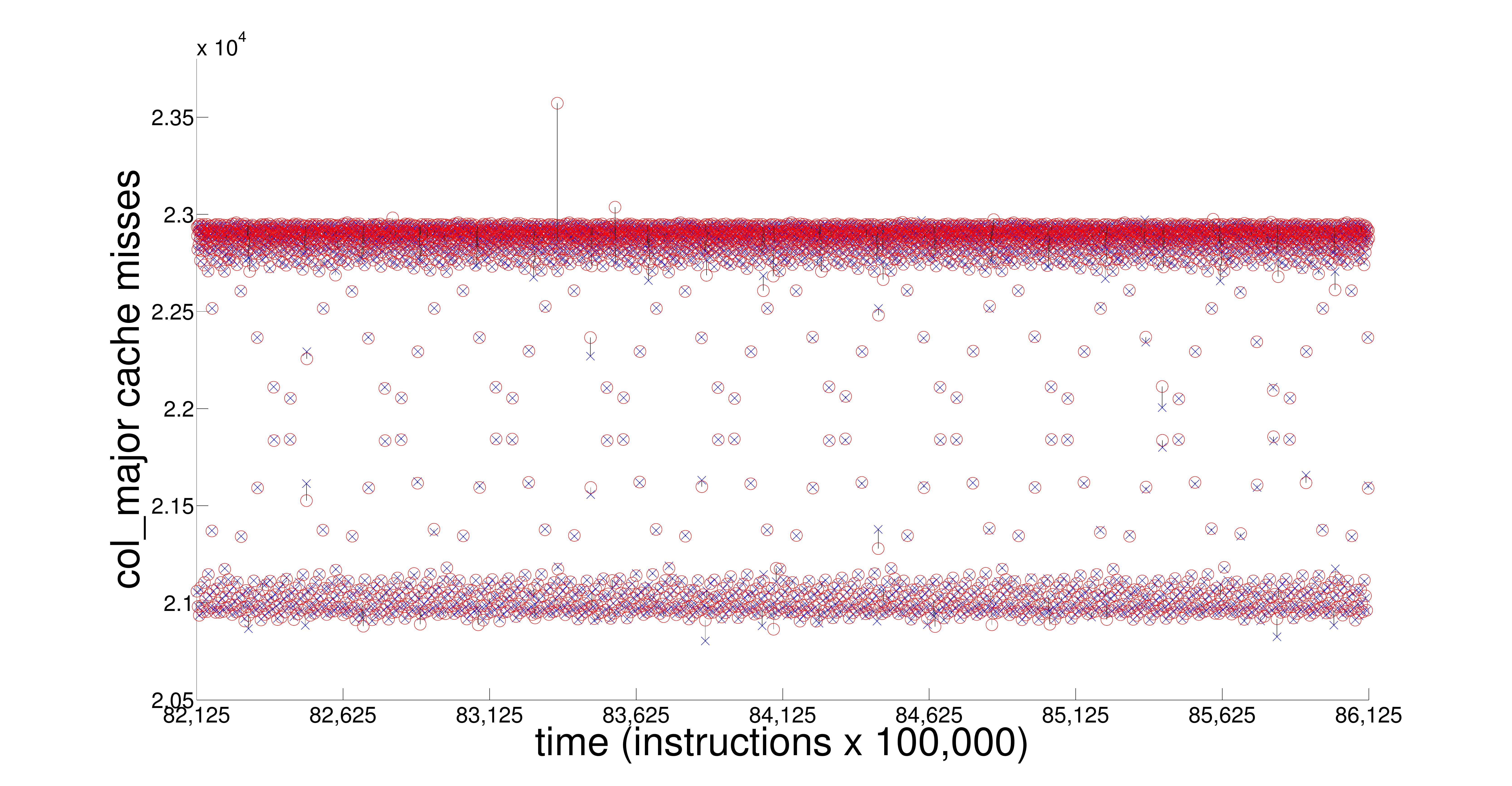}
    \caption{A forecast of the last 4,000 points of the signal in
      Figure~\ref{fig:cache} using an LMA-based strategy on the
      embedding in Figure~\ref{fig:embedding}.  Red circles and blue
      $\times$s are the true and predicted values, respectively;
      vertical bars show where these values differ. }
\label{fig:cachePredTS}
\end{figure}

\begin{figure}[htbp]
  \centering
    \includegraphics[width=0.5\textwidth]{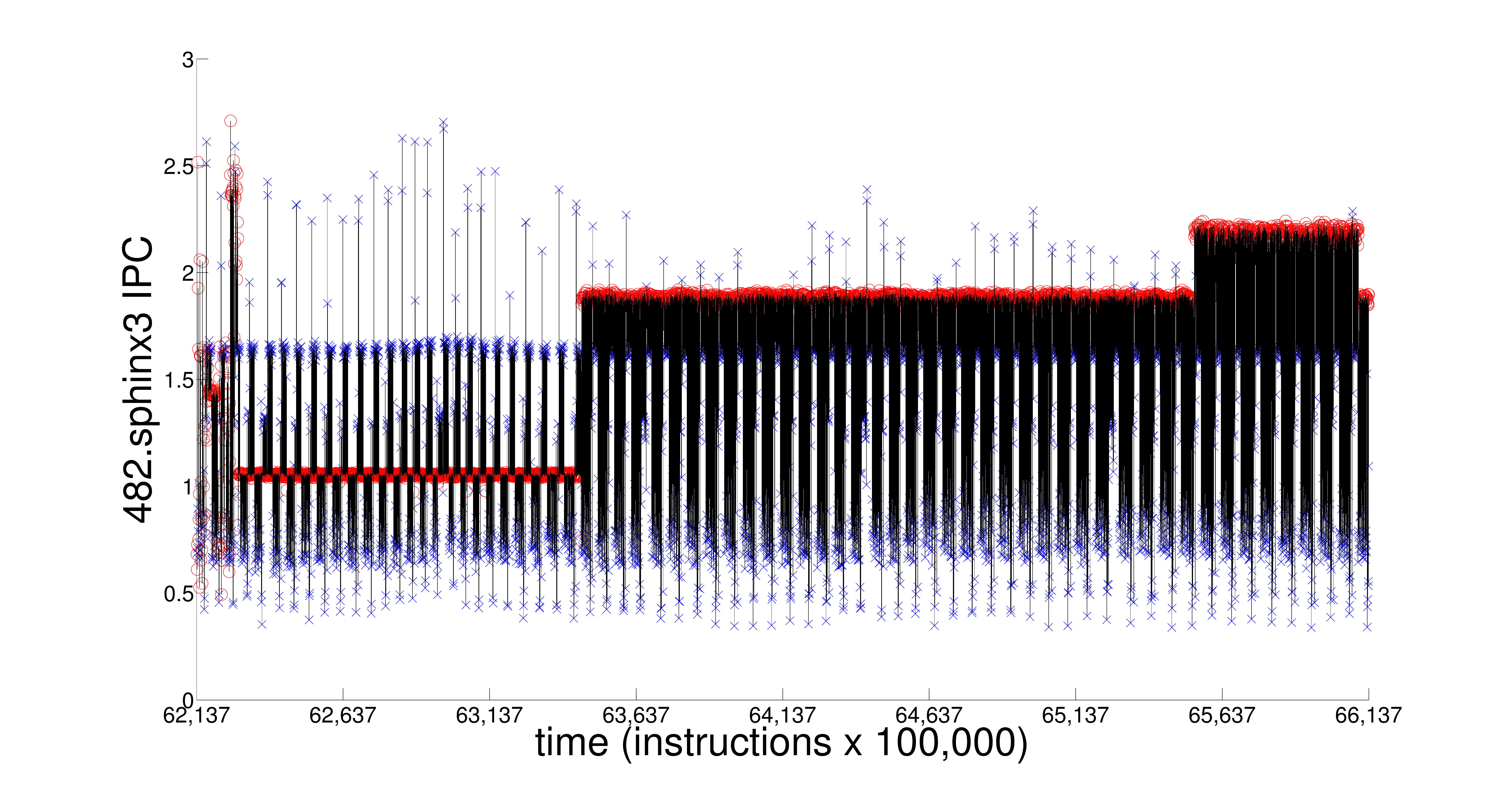}
     \caption{An LMA-based forecast of the last 4,000 points of a
       processor-load performance trace from the {\tt 482.sphinx3}
       benchmark.  Red circles and blue $\times$s are the true and
       predicted values, respectively; vertical bars show where these
       values differ.}
\label{fig:predsphinx}
\end{figure}

Given a nonlinear model of a deterministic dynamical system in the
form of a delay-coordinate embedding like Figure~\ref{fig:embedding},
one can build deterministic forecast algorithms by capturing and
exploiting the geometry of the embedding.  Many techniques have been
developed by the dynamical systems community for this purpose
(e.g.,~\cite{casdagli-eubank92,weigend-book}).  Perhaps the most straightforward
is the ``Lorenz method of analogues'' (LMA), which is essentially
nearest-neighbor prediction in the embedded state
space~\cite{lorenz-analogues}.  Even this simple algorithm---which
builds predictions by finding the nearest neighbor in the embedded
space of the given point, then taking that neighbor's path as the
forecast---works quite well on the trace in Figure~\ref{fig:cache}, as
shown in Figure~\ref{fig:cachePredTS}.
On the other hand, if we use the same approach to forecast the
processor load\footnote{Instructions per cycle, or IPC} of the {\tt
  482.sphinx3} program from the SPEC cpu2006 benchmark suite, running
on an Intel i7\textsuperscript{\textregistered}-based machine, the
prediction is far less accurate; see Figure~\ref{fig:predsphinx}.

Table~\ref{tab:PredError} presents detailed results about the
prediction accuracy of this algorithm on four different examples: the
{\tt col\_major} and {\tt 482.sphinx3} programs in
Figures~\ref{fig:cachePredTS} and~\ref{fig:predsphinx}, as well as
another simple microkernel that initializes the same matrix as {\tt
  col\_major}, but in row-major order, and another complex program
({\tt 403.gcc}) from the SPEC cpu2006 benchmark suite.  Both
microkernels were run on the Intel Core
Duo\textsuperscript{\textregistered} machine; both SPEC benchmarks
were run on the Intel i7\textsuperscript{\textregistered} machine.  We
calculated a figure of merit for each prediction as follows.  We held
back the last $k$ elements\footnote{Several different prediction
  horizons were analyzed in our experiment; the results reported in
  this paper are for $k$=4000} of the $N$ points in each measured time
series, built the forecast model by embedding the first $N-k$ points,
used that embedding and the LMA method to predict the next $k$ points,
then computed the Root Mean Squared Error (RMSE) between the true and
predicted signals:
$$RMSE = \sqrt{\frac{\sum_{i=1}^k(c_i-\hat{p_i})^2}{k}}$$
To compare the success of predictions across signals with different
units, we normalized RMSE as follows:
$$nRMSE = \frac{RMSE}{X_{max,obs}-X_{min,obs}}$$
%
% The smaller the nRMSE, obviously, the more accurate the prediction.

  \begin{table}[htbp]
   \renewcommand{\arraystretch}{1.3}
   \caption{Normalized root mean squared error (nRMSE) of 4000-point
     predictions of memory \& processor performance from different
     programs.}
   \label{tab:PredError}
   \centering
   % Some packages, such as MDW tools, offer better commands for making tables
   % than the plain LaTeX2e tabular which is used here.
%   \begin{tabular}{|c|c|c|c|c|c|}
%     \hline
%      & cache misses NRMSE & IPC NRMSE \\
%     \hline
%     \tt{row\_major}  & 0.0102  & 0.0095 \\
%          \hline
%     \tt{col\_major}  &0.0055& 0.0071 \\
%     \hline
%     \tt{403.gcc}  & 0.0865& 0.1805 \\
%          \hline
%     \tt{482.sphinx3} & 0.1142& 0.2946 \\
%      \hline
%   \end{tabular}
% \end{table}
   \begin{tabular}{|c|c|c|c|c|c|}
     \hline
      & cache miss rate & instrs per cycle \\
     \hline
     \tt{row\_major}  & 0.0324  & 0.0778 \\
          \hline
     \tt{col\_major}  &0.0080& 0.0161 \\
     \hline
     \tt{403.gcc}  & 0.1416& 0.2033 \\
          \hline
     \tt{482.sphinx3} & 0.2032& 0.3670 \\
      \hline
   \end{tabular}
 \end{table}

The results in Table~\ref{tab:PredError} show a clear distinction
between the two microkernels, whose future behavior can be
predicted effectively using this simple deterministic modeling
strategy, and the more-complex SPEC benchmarks, for which this
prediction strategy does not work nearly as well.
% Removed for space
% For both processor load (IPC) and memory usage (cache-miss rate),
% forecasts for {\tt 482.sphinx3} and {\tt 403.gcc} are much worse than
% for \verb|col_major| or \verb|row_major|.
%
This begs the question: If these traces all come from deterministic
systems---computers---then why are they not equally predictable?  Our
conjecture is that the sheer complexity of the dynamics of the SPEC
benchmarks running on the Intel i7\textsuperscript{\textregistered}
machine make them effectively impossible to predict.

%% [[If space: Add a segue sentence: next section uses permutation
%% entropy to explore that conjecture.]]

\section{Measuring Complexity}\label{sec:meaComplex}

For the purposes of this paper, one can view entropy as a measure of
complexity and predictability in a time series.  A high-entropy time
series is almost completely unpredictable---and conversely.  This can
be made more rigorous: Pesin's relation \cite{pesin77} states that in
chaotic dynamical systems, the Shannon entropy rate is equal to the
sum of the positive Lyapunov exponents, $\lambda_i$. The Lyapunov
exponents directly quantify the rate at which nearby states of the
system will diverge with time: $\left| \Delta x(t) \right| \approx
e^{\lambda t} \left| \Delta x(0) \right|$.  The faster the divergence,
the more difficult prediction becomes.

Utilizing entropy as a measure of temporal complexity is by no means a new idea
\cite{Shannon1951, mantegna1994linguistic}.  Its effective usage
requires categorical data: $x_t \in \mathcal{S}$ for some finite or
countably infinite \emph{alphabet} $\mathcal{S}$, and data taken from
real-world systems is effectively real-valued.  To get around this,
one must discretize the data---typically by binning.  Unfortunately,
this is rarely a good solution to the problem, as the binning of the
values introduces an additional dynamic on top of the intrinsic
dynamics whose entropy is desired.  The field of symbolic dynamics
studies how to discretize a time series in such a way that the
intrinsic behavior is not perverted, but these methods are fragile in
the face of noise and require further understanding of the underlying
system, which defeats the purpose of measuring the entropy in the
first place.

Bandt and Pompe introduced \emph{permutation entropy} (PE)
% also called permutation complexity
as a ``natural complexity measure for time series"
\cite{bandt2002per}.  Permutation entropy employs a method of
discretizing real-valued time series that follows the intrinsic
behavior of the system under examination.  Rather than looking at the
statistics of sequences of values, as is done when computing the
Shannon entropy, permutation entropy looks at the statistics of the
\emph{orderings} of sequences of values using ordinal analysis.
Ordinal analysis of a time series is the process of mapping successive
time-ordered elements of a time series to their value-ordered
permutation of the same size.  By way of example, if $(x_1, x_2, x_3)
= (9, 1, 7)$ then its \emph{ordinal pattern}, $\phi(x_1, x_2, x_3)$,
is $231$ since $x_2 \leq x_3 \leq x_1$.  This method has many
features; among other things, it is robust to noise and requires no
knowledge of the underlying mechanisms.

% It has been shown [amigo?] that the
% normalized permutation entropy \ref{eq:permEntropy} is equal to the
% Kolmogorov-Sinai entropy and therefore is a proxy for the rate at
% which a system becomes unpredictable.

% This spectrum is outlined very nicely in \cite{amigo2012permutation}:
% ``Periodic or quasiperiodic sequences have vanishing or negligible
% complexity. At the opposite end, independent and identically
% distributed random sequences (white noise) have asymptotically
% divergent permutation entropies, owing to the fact that the number of
% allowed (or ``admissible'') ordinal patterns grows superexponentially
% with length. Between both ends lie the kind of sequences we are
% interested in.''

\begin{mydef}[Permutation Entropy]
  Given a time series $\{x_t\}_{t = 1,\dots,T}$. Define
  $\mathcal{S}_n$ as all $n!$ permutations $\pi$ of order $n$. For
  each $\pi \in \mathcal{S}_n$ we determine the relative frequency of
  that permutation occurring in $\{x_t\}_{t = 1,\dots,T}$: $$ p(\pi) =
  \frac{\left|\{t|t\le T-n,\phi(x_{t+1},\dots,x_{t+n}) =
      \pi\}\right|}{T-n+1}$$ Where $|\cdot|$ is set cardinality. The
  \emph{permutation entropy} of order $n\ge 2$ is defined
  as$$H(n) = - \sum_{\pi \in
      \mathcal{S}_n} p(\pi)\log_2p(\pi)$$
\end{mydef}
Notice that $0\le H(n) \le \log_2(n!)$ \cite{bandt2002per}.  With this
in mind, it is common in the literature to normalize permutation
entropy as follows: $\frac{H(n)}{\log_2(n!)}$.  With this convention,
``low'' entropy is close to 0 and ``high'' entropy is close to 1.
Finally, it should be noted that the permutation entropy has been
shown to be identical to the Shannon entropy for many large classes of
systems \cite{amigo2012permutation}.

In practice, calculating permutation entropy involves choosing a good
value for the wordlength $n$. The key consideration
here is that the value be large enough that forbidden ordinals are
discovered, yet small enough that reasonable statistics over the
ordinals are gathered: e.g., $\displaystyle n =
\operatornamewithlimits{argmax}_\ell\{T > 100 \ell!\}$, assuming an
average of 100 counts per ordinal. In the literature, $3\le n \le 6$
is a standard choice---generally without any formal justification. In
theory, the permutation entropy should reach an asymptote with
increasing $n$, but that requires an arbitrarily long time series. In
practice, what one should do is calculate the \emph{persistent}
permutation entropy by increasing $n$ until the result converges, but
data length issues can intrude before that convergence is reached.

% Tables~\ref{tab:TCMpe} \& \ref{tab:IPCpe}
Table~\ref{tab:pe} shows the permutation entropy results for the
examples considered in this paper, with the nRMSPE prediction
accuracies from the previous section included alongside for easy
comparison.
  \begin{table}[htbp]
   \caption{Prediction error (in nRMSPE) and permutation entropy (for
     different wordlengths $n$}
   \label{tab:pe}
   \centering
   \begin{tabular}{|c|c|c|c|c|c|}
     \hline
      {\bf cache misses} & error   &$n=4$&$n=5$&$n=6$\\
     \hline
     \tt{row\_major}  & 0.0324  &0.6751&0.5458&0.4491\\
     \hline
     \tt{col\_major}  & 0.0080  &0.5029&0.4515&0.3955 \\
     \hline
     \tt{403.gcc}  & 0.1416  &0.9916&0.9880&0.9835\\
     \hline
     \tt{482.sphinx3}  & 0.2032  &0.9913&0.9866&0.9802 \\
      \hline
      \hline
      {\bf insts per cyc} & error   &$n=4$&$n=5$&$n=6$\\
     \hline
      \tt{row\_major} & 0.0778  &0.9723&0.9354&0.8876\\
     \hline
      \tt{col\_major} & 0.0161   &0.8356&0.7601&0.6880\\
     \hline
     \tt{403.gcc} & 0.2033   &0.9862&0.9814&0.9764\\
     \hline
     \tt{482.sphinx3} & 0.3670 & 0.9951&0.9914&0.9849 \\
      \hline
   \end{tabular}
 \end{table}
The relationship between prediction accuracy and the permutation
entropy (PE) is as we conjectured: performance traces with high
PE---those whose temporal complexity is high, in the sense that little
information is being propagated forward in time---are indeed harder to
predict using the simple deterministic forecast model described in the
previous section.  The effects of changing $n$ are also interesting:
using a longer wordlength generally lowers the PE---a natural
consequence of finite-length data---but the falloff is less rapid in
some traces than in others, suggesting that those values are closer to
the theoretical asymptote that exists for perfect data.  The
persistent PE values of 0.5--0.6 for the {\tt row\_major} and {\tt
  col\_major} cache-miss traces are consistent with dynamical chaos,
further corroborating the results of~\cite{mytkowicz09}.  (PE values
above 0.97 are consistent with white noise.)  Interestingly, the
processor-load traces for these two microkernels exhibit more temporal
complexity than the cache-miss traces.  This may be a consequence of
the lower baseline value of this time series.

% In Table~\ref{tab:TCMpe} \& \ref{tab:IPCpe} the behavior of the PEs as
% they limit to their asymptotic values is observed. A feature of note
% is that some of the signals drop off and some do not.  For example at
% $m=4$ {\tt col\_major} is at 0.8356, a fairly high entropy, however as
% $m$ is increased to 6 the PE plummets to 0.6880: a value consistent
% with a chaotic time series. Notice this also corresponds to a very low
% nRMSE. In contrast, consider the permutation entropy of the processor
% load of {\tt 482.sphinx3}: with $m=4$ it is 0.9951 and only drops to
% 0.9849 at $m=6$. This is consistent with the signal also having the
% highest nRMSE.

\section{ Conclusions \& Future Work}\label{sec:conc}

The results presented here suggest that permutation entropy---a
ordinal calculation of forward information transfer in a time
series---is an effective metric for predictability of computer
performance traces. Experimentally, traces with a persistent PE
$\gtrapprox 0.97$ have a natural level of complexity that may
overshadow the inherent determinism in the system dynamics, whereas
traces with PE $\lessapprox 0.7$ seem to be highly predictable (viz.,
at least an order of magnitude improvement in nRMSPE).

If information is the limit, then gathering and using more information
is an obvious next step.  There is an equally obvious tension here
between data length and prediction speed: a forecast that requires
half a second to compute is not useful for the purposes of real-time
control of a computer system with a MHz clock rate.  Another
alternative is to sample several system variables simultaneously and
build multivariate delay-coordinate embeddings.  Existing approaches
to that are computationally prohibitive
\cite{cao-multivariate-embedding}.  We are working on alternative
methods that sidestep that complexity.

\section*{Acknowledgment}
This work was partially supported by NSF grant \#CMMI-1245947 and ARO
grant \#W911NF-12-1-0288.

\bibliographystyle{IEEEtran}
\bibliography{IEEEabrv,bibliofile}

\end{document}